\documentclass[conference]{IEEEtran}
\IEEEoverridecommandlockouts

\usepackage{cite}
\usepackage{amsmath,amssymb,amsfonts}
\usepackage{algorithmic}
\usepackage{algorithm}
\usepackage{graphicx}
\graphicspath{{../figures/}{figures/}{./}}
\usepackage{textcomp}
\usepackage{authblk}
\usepackage{xcolor}
\def\BibTeX{{\rm B\kern-.05em{\sc i\kern-.025em b}\kern-.08em
    T\kern-.1667em\lower.7ex\hbox{E}\kern-.125emX}}

\usepackage{booktabs}
\usepackage{array}
\usepackage{multirow}
\usepackage{url}

\newcommand{\auc}{AUC}

\begin{document}

\title{Towards Multilingual End-of-Turn Detection using Frozen Speech Encoders}

\author{Rini Sharon$^{\star}$ \qquad Kadri Hacioglu$^{\star}$ \qquad Andreas Stolcke$^{\star}$}

\affil{$^{\star}$ Uniphore} 




\maketitle

\begin{abstract}
A spoken dialogue system must judge when the user has yielded the floor. This end-of-turn (EoT) detector is normally trained and tuned afresh for each language, which is costly to scale to new languages. We instead build one on a frozen multilingual encoder, training only a small fixed-size head. Using five CALLHOME languages, we investigate three questions. First, how well a head trained on one language transfers zero-shot to the others. Second, how much target-language adaptation is required to recover full task performance. Third, whether a single fixed decision threshold holds across languages without per-language tuning. We find that the head transfers well, even between tonal and pitch-accent languages, and that a few hours of audio recover nearly all the loss, matching in-language training. A frozen multilingual encoder with one lightweight head is a practical, language-agnostic baseline for EoT detection.
\end{abstract}

\begin{IEEEkeywords}
End-of-turn detection, turn-taking, cross-lingual transfer, streaming ASR,
frozen speech encoders, latency, conversational AI.
\end{IEEEkeywords}

\section{Introduction}
Conversation runs in turns, with speakers following one another with little gap or
overlap~\cite{skantze2021review}, and a spoken dialogue system must do the same. Before it
can reply, it has to detect the end of the user's turn (EoT) as the audio arrives, in real
time and with no access to what comes next. In speech systems this task is also called
endpointing, and it is a core component of voice assistants and other dialogue systems. The
timing of the decision matters, because a late decision leaves an unnatural pause while an
early one interrupts the user~\cite{levinson2015timing}.


The simplest detector waits for a fixed span of silence, but silence is a weak cue, since pauses within a turn are common and can last longer than the gaps between turns~\cite{heldner2010}. Speakers instead mark the end of a turn earlier, through prosodic, syntactic, and semantic cues that a listener perceives before the final silence~\cite{sacks1974systematics,ward2000prosodic,gravano2011turn,wightman1992segmental,deruiter2006projecting}. Modern EoT detectors read these cues to respond faster and more reliably than a silence timeout~\cite{ferrer2002speaker,ferrer2003prosody,arsikere2014enhanced,chang2017endpoint,li2020streaming}, usually with a frame-level classifier over the features of a neural speech encoder~\cite{chang2022turn,shangguan2021dissecting,bijwadia2022unified}. Recent detectors add real-time refinements, from anticipating the endpoint before it arrives to segmenting the primary speaker in multi-talker audio~\cite{udupa2026endpoint,helwani2026hierarchical}. Such a detector is judged on two costs that trade off. One is latency, the delay between the true end of the turn and the moment the detector fires. The other is the false-alarm rate, how often it fires while the user is still speaking. Firing sooner lowers latency but raises false alarms, so a system settles on an operating point that balances the two.

These detectors increasingly have to work in many languages, yet turn-taking has been studied
far less across languages than within one~\cite{skantze2021review}. Detectors are still built
and tuned one language at a time~\cite{aldeneh2018endofturn,chen2025fullduplex}, and how much of
a detector carries over when the language changes is not well understood. Building one for a new
language is expensive in two ways. It needs labeled turn boundaries in that language, which are
costly to annotate, and it needs a decision threshold, which production systems set separately
for each language by sweeping until the false-alarm rate meets a target
budget~\cite{pipecat2025smartturn,livekit2024turn}. Repeating both for every new language scales
poorly, so we ask whether it is necessary.


Cross-lingual EoT has had little direct study, and most of it comes from ASR systems. There, English-trained embeddings transfer to new languages with light per-language adaptation~\cite{maas2018eou}, the turn end is often recast as semantic endpointing over the recognized transcript~\cite{assemblyai2025endpointing,popit2025thai}, and recent systems use conversational context to decide the same boundary~\cite{speechmatics2025semantic}. Production streaming recognizers now ship multilingual endpointing directly~\cite{assemblyai2025multilingual}, and recent production endpointers report large reductions in false interruptions across many languages~\cite{livekit2025intl} and favor one multilingual model over per-language ones~\cite{livekit2026v1}. All of these work through the recognizer and its text, leaving open whether the acoustic turn-taking signal transfers across languages on its own. The two works closest to ours model that signal directly, JAL-Turn~\cite{jalturn2026} and multilingual voice activity projection~\cite{inoue2024multilingual}, both larger trained models that we contrast with our lighter single-head setup below.

JAL-Turn is a turn-taking model built for full-duplex agents. It fuses two frozen encoders, a multilingual ASR encoder for
linguistic and semantic cues and a separate one for low-level acoustics, through a trained
cross-attention Transformer, and predicts hold-versus-shift states from one model trained jointly
across many languages. Where it shows that a shared encoder can drive accurate multilingual
turn-taking, we ask the prior question of how far a single head reaches on its own, trained on one
language and tested on another it has never seen.

Voice activity projection (VAP), the other, predicts the upcoming activity of both speakers
continuously, rather than detecting a single
boundary~\cite{ekstedt2022vap,ekstedt2022prosody,inoue2024realtime,inoue2025backchannel},
building on a line of continuous neural turn-taking
models~\cite{skantze2017continuous,roddy2018investigating,ekstedt2020turngpt}. Trained on English,
Mandarin, and Japanese, a monolingual VAP transfers poorly to the others, while a single model
trained jointly on all three matches the monolingual baselines~\cite{inoue2024multilingual}. This
fits the finding that between-turn timing is largely shared across languages, which gives a model
similar dynamics to learn in each~\cite{stivers2009universals,levinson2015timing}. Our setup
differs from VAP in three ways that matter for a cross-lingual claim. VAP retrains its encoder and
projection network for each language, so its
cross-lingual results mix representation with retraining, while ours come from the representation
alone. It is also scored offline on recorded dialogue, like related turn-taking
models~\cite{wang2024fusion}, which can overstate what a live system sees, whereas we score our head
causally and report the latency and false-alarm rate a live system faces. Finally, because we run at
a fixed operating point, we can ask whether one decision threshold holds across languages unchanged,
a question that does not arise offline.

The natural place to look for shared structure is the encoder. Multilingual encoders such as
wav2vec~2.0, XLS-R, MMS, and Whisper learn speech representations from large multilingual
corpora~\cite{baevski2020wav2vec2,babu2022xlsr,pratap2024mms,radford2023whisper}, and a single
shared encoder transfers across languages, sometimes matching or surpassing per-language training
on lower-resource languages~\cite{tan2020crosslingual,conneau2021xlsr}. Such encoders are often
used frozen, as fixed feature extractors read by a small trained
head~\cite{alain2017probing,yang2021superb,shi2023mlsuperb,hsu2021hubert,chen2022wavlm}. Recent
work probes which turn-taking cues these representations carry, finding they draw on both prosody
and lexical content~\cite{russell2026prosodic}.

We take this route for EoT, where prior work has left three things untested. Cross-lingual EoT has
been tested only through the recognizer and its text, the models that read the acoustic signal
directly are trained per language, scored offline, or built at scale, and their decision threshold
is fixed rather than reused across languages. These motivate our three questions, studied on a
single frozen encoder with one small head, which together form our contributions.

\begin{itemize}
\item \textbf{(RQ1) Does a head trained on one language work on another?} We measure
zero-shot cross-lingual transfer of the acoustic signal alone, with both a full-context and a
causal encoder across all five languages (Section~\ref{sec:e1}).

\item \textbf{(RQ2) How much target-language audio recovers the loss?} We measure
how quickly the remaining EoT detection gap closes as target audio is added, and ask whether
onboarding a language is light adaptation or a full rebuild
(Section~\ref{sec:e2}).

\item \textbf{(RQ3) Does a threshold calibrated on one language hold on another?}
We test whether a decision threshold keeps its false-alarm rate and latency across
languages, since a real-time system runs at a single operating point and is judged there
(Section~\ref{sec:e3}).
\end{itemize}


We test this on five languages from the CALLHOME telephone corpus~\cite{ldc1997callhome} that
span three language families~\cite{glottolog2023}, so we can ask whether transfer depends on how
related the languages are, not only on how they use pitch. Mandarin is tonal, where pitch is
lexical and distinguishes word meaning~\cite{yip2002tone}, and Japanese is pitch-accent, where the
position of an accent within a word can change its meaning~\cite{beckman1986japanese}, while
English, German, and Spanish use pitch post-lexically for phrasing~\cite{ladd2008intonational}.
Since turn-final cues are largely prosodic and pitch is a major part of
them~\cite{gravano2011turn,bogels2015intonation}, a language that already spends pitch on word
meaning may have less of it free to mark a turn end, which makes Mandarin and Japanese the most
demanding test of transfer.


\section{Dataset and Method}
\label{sec:data}
We want a comparison in which any cross-language difference is attributable to the representation
rather than to the data or the head. We therefore hold everything but the language fixed, the
corpus and its domain, the labeling rule, the encoder, the head, and the training budget, so that
language is the only variable that changes across conditions.

\subsection{Data and labels}
We use the CALLHOME corpus of unscripted two-party telephone
conversations~\cite{ldc1997callhome}, taking its English, Spanish, German, Mandarin, and Japanese
partitions. All five share the same domain, the same $8$\,kHz telephone channel, and the same
recording protocol. The partitions range from $120$ to $140$ conversations, with $140$ each for
English, Spanish, and Mandarin and $120$ each for German and Japanese, for $660$ conversations and
about $99$\,h of raw audio in total. We decode the audio to $16$\,kHz mono for feature extraction.
After applying the EoT rule below, this yields $74{,}646$ labeled examples spanning $59.5$\,h.

\textbf{From CALLHOME timing to EoT labels.} CALLHOME provides time-stamped speech segments with
speaker labels but no turn-end marks. We derive turn ends from this timing with
Algorithm~\ref{alg:eot}, applied identically to every language. It merges consecutive same-speaker
segments into turns and labels each turn end.


\begin{algorithm}[t]
\caption{EoT labeling for one conversation.}
\label{alg:eot}
\small
\begin{algorithmic}[1]
\REQUIRE diarized (speaker, start, end) segments, sorted by start time
\REQUIRE merge gap $T_m{=}0.5$\,s, trailing context $T_e{=}1.3$\,s, overlap tolerance $T_o{=}0.05$\,s
\STATE Merge consecutive same-speaker segments $\le T_m$ apart, giving turns with start $\mathrm{st}_i$ and end $e_i$
\FOR{each turn $i$}
  \STATE $g \gets (\text{onset of next speech}) - e_i$ \COMMENT{gap to next talker, negative if they overlap}
  \STATE $\mathit{kind}\gets$ \emph{final} if no speech follows, \emph{hold} if the same speaker resumes, \emph{shift} if $g \ge -T_o$, else skip this turn (active overlap)
  \STATE $\tau \gets \textsc{VadSnap}(e_i)$ \COMMENT{nearest Silero speech offset within $\pm0.4$\,s}
  \STATE emit example over $[\,\mathrm{st}_i,\ \min(\tau + T_e,\ \text{next onset})\,]$, boundary $\tau$, kind $\mathit{kind}$
  \STATE label its frames with the collar, Eq.~\eqref{eq:collar}
\ENDFOR
\end{algorithmic}
\end{algorithm}

The three boundary kinds are all genuine turn ends, and the merge step is what makes a \emph{hold}
one. Short same-speaker pauses such as hesitations or breaths, at most $T_m$, are folded into the
turn, so a surviving hold is a completed utterance whose boundary still carries turn-final prosody. We skip
active overlaps, where the floor is not cleanly yielded and the turn end is ill-defined. Snapping the
boundary to the nearest Silero VAD offset~\cite{silero2021vad} places the label at the acoustic end
of speech rather than at the annotated segment boundary.

\begin{figure}[t]
\centering
\includegraphics[width=\columnwidth]{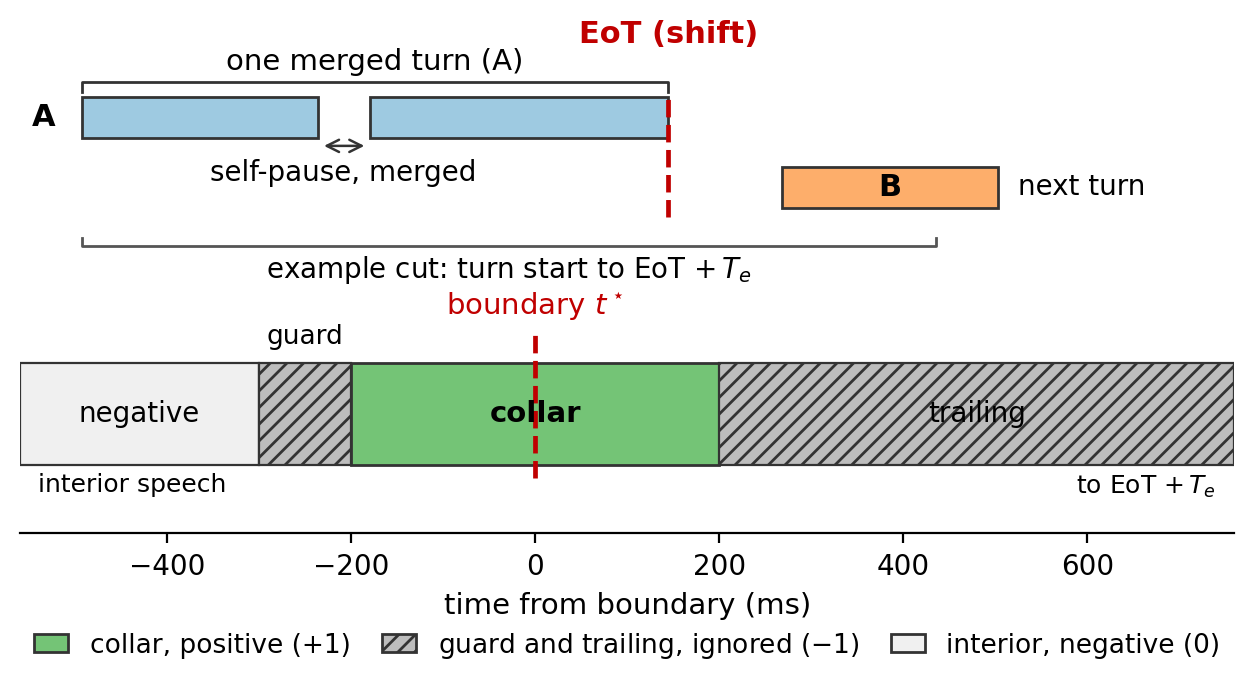}
\caption{How one conversation becomes labeled examples, shown for a \emph{shift} boundary, with the
frame-level target zoomed in below.}
\label{fig:label}
\end{figure}

\textbf{Frame-level labels.} Because the detector scores every frame rather than each turn once, we
attach a label to each frame. Writing $t^\star$ for the frame at the EoT boundary, we treat a small
window around it as positive, the \emph{collar}, with half-width $W$. The label of frame $j$ is
\begin{equation}
\label{eq:collar}
y_j =
\begin{cases}
\phantom{-}1 & \text{if } |\,j-t^\star\,| \le W,\\
\phantom{-}0 & \text{if } j < t^\star - G,\\
-1 & \text{otherwise,}
\end{cases}
\end{equation}
with guard $G{=}W{+}5$. Frames inside the $\pm W$ collar are positive, since any of them is an
acceptable moment to fire, and using a window rather than a single frame keeps the target robust to
small timing errors. Every frame more than $G$ before the boundary is negative, so a detector that
fires mid-turn, including on a within-turn pause, counts as a false alarm. The guard band just
before the collar and the trailing context just after the boundary are left unscored and marked
$-1$ (Figure~\ref{fig:label}). We use $W{=}10$ ($200$\,ms) by default.

Two properties of this scheme matter for scoring. Because the trailing frames are unscored, the
negatives are interior speech rather than silence, so the score cannot be inflated by trivially easy
silence frames, and a chance detector sits at $0.5$. The positive frames are a minority, which
class-balanced training equalizes against the negatives at fit time.


\subsection{Comparability controls}
\textbf{Label parity across languages.} Applying one rule to every language is only a fair
comparison if it produces similar label distributions across them. If one language had far more
positive frames, or a very different mix of boundary types, a measured transfer gap could reflect
that imbalance rather than the acoustics we care about. We therefore inspect the per-language label
statistics before running any experiment (Table~\ref{tab:parity}), and two quantities matter. The
first is the positive-frame fraction, the share of an example's frames that fall inside the EoT
collar and so count as positive. Across the five languages it ranges from $0.206$ for Spanish to
$0.302$ for Mandarin, a spread of under $0.10$, which we take as our parity tolerance. The second is
the boundary-kind mix, the relative frequency of shifts, holds, and finals. Shifts dominate in every
language, same-speaker holds make up most of the remainder, and final turns are negligible at around
$0.01$. Spanish is the extreme on both counts, with the fewest holds and the lowest positive-frame
fraction, yet it still falls within tolerance, so no language is labeled differently enough to
distort the transfer comparison.

\begin{table}[t]
\centering
\caption{Per-language label statistics under the shared EoT rule.}
\label{tab:parity}
\begin{tabular}{lrcccc}
\toprule
Language & Examples & Pos.\ frac. & Shift & Hold & Final \\
\midrule
English  & 16{,}454 & 0.272 & 0.683 & 0.308 & 0.009 \\
Spanish  & 12{,}102 & 0.206 & 0.828 & 0.160 & 0.012 \\
German   & 12{,}386 & 0.233 & 0.790 & 0.201 & 0.010 \\
Mandarin & 18{,}762 & 0.302 & 0.717 & 0.275 & 0.007 \\
Japanese & 14{,}942 & 0.243 & 0.759 & 0.233 & 0.008 \\
\bottomrule
\end{tabular}
\end{table}

\textbf{Data splits.} Within each language we split the data three ways by duration, $50\%$ to fit
the head, $25\%$ to make every threshold and model choice, and a held-out $25\%$ that produces all
reported numbers. The split is speaker-disjoint, so no talker appears in more than one part and a
head cannot earn its score by recognizing a familiar voice instead of learning where turns end. We
use the same partition in every experiment.


\subsection{Model}
\textbf{Frozen encoder.} We use the Whisper-small encoder as a fixed feature extractor, never
fine-tuned. It produces $768$-dimensional features at $50$\,Hz, one frame every $20$\,ms, and we
choose the read-out layer by a per-layer sweep on the source data~\cite{radford2023whisper}. We read
it in two modes that differ only in how much audio each frame may use. The \emph{oracle} mode makes
one bidirectional pass over the whole example, giving every frame unbounded lookahead. It cannot be
deployed and serves only as a full-context reference ceiling, as streaming models are usually placed
against their full-context counterparts~\cite{li2020streaming,shi2021emformer}. The \emph{causal}
mode is streaming-safe, processing the signal in $160$\,ms chunks and recomputing, for each chunk,
the log-mel input from the audio prefix rather than the whole file. This closes the two paths by
which future audio would leak in, the convolutional front end's finite receptive field and Whisper's
global log-mel normalization, so a frame's features depend only on audio up to its chunk boundary
and stay bit-for-bit unchanged as later audio arrives. A fixed $10$\,ms lookahead (half an output
frame) remains in the front end, which we document rather than remove. The causal mode is the one
behind every latency claim, and we call the oracle-to-causal gap the cost of causality.

\textbf{Head.} The head is a small $L_2$-regularized logistic probe on standardized features, with
the same capacity and training budget in every language, so it adds no language-specific capacity.
Its one hyperparameter, the $L_2$ strength, is tuned on the source language alone with speaker-split
folds, never on the target. A single-hidden-layer probe later serves as a nonlinear robustness
check.

\subsection{Metrics and statistics}
\textbf{AUC.} Our primary metric is \auc{}, the area under the ROC curve~\cite{fawcett2006roc}. It
is the probability that the head scores a randomly chosen collar frame above a randomly chosen
interior frame, so it measures how well the head ranks turn-end frames above ongoing speech. It is
threshold-free, with chance at $0.5$, and we report it with confidence intervals.

\textbf{Latency and false alarms.} A real-time system is judged on latency and false-alarm rate at a
fixed operating point, so we report both. A \emph{trigger} turns the per-frame scores into a single
endpoint decision. It smooths the scores with an exponential moving average ($\alpha{=}0.3$) and
fires once the smoothed score has stayed above the threshold $\theta$ for three consecutive frames,
matching our production streaming evaluator. Latency is the signed gap between that firing and the
true boundary, measured at $20$\,ms resolution and negative when the detector fires early. The
threshold $\theta$ is the operating point, set on a development split to hold false alarms at a
target budget, here $5\%$, and we set it in two ways. An \emph{in-language} threshold is chosen on
the same language it is evaluated on, an optimistic in-sample ceiling, whereas a \emph{transferable}
threshold is calibrated on one language and applied unchanged to another, the realistic
cross-lingual case. Being threshold-free, \auc{} is unaffected by this distinction.

\textbf{Transfer gap.} Our headline cross-lingual measure is the transfer gap $\Delta$. Let
$A_{st}$ be the \auc{} of a head trained on language $s$ and evaluated on language $t$, over a
language set $L$. The transfer gap is the mean same-language score, on the diagonal of the transfer
matrix, minus the mean cross-language score, off the diagonal,
\begin{equation}
\label{eq:gap}
\Delta = \frac{1}{|L|}\sum_{\ell \in L} A_{\ell\ell}
\;-\; \frac{1}{|L|\,(|L|-1)} \sum_{\substack{s,t \in L \\ s \neq t}} A_{st}.
\end{equation}
A small $\Delta$ means a head ranks turn ends on other languages almost as well as on the one it
was trained on, which is what we mean by language-universal transfer.

\textbf{Statistical testing.} All confidence intervals come from a speaker-block bootstrap. We
resample whole speaker blocks with replacement, recompute the statistic on each resample, and take
its $2.5$ and $97.5$ percentiles, so the intervals reflect variation across talkers rather than
across individual frames~\cite{efron1986bootstrap,bisani2004bootstrap}. To compare two conditions,
such as the diagonal and the off-diagonal of the transfer matrix, we use the same resampled blocks
for both, a paired bootstrap, and read significance from the fraction of resamples in which the
difference changes sign~\cite{koehn2004significance}.

\section{Experiments and Results}
Each research question looks at a different per-language cost, and we take them in the order a new
language runs into them, first training a head (RQ1), then adapting it (RQ2), and finally setting
its threshold (RQ3). In every experiment we run the same head under both encoders of
Section~\ref{sec:data}, the causal encoder, which a real-time system can actually use and which
produces all of our latency and false-alarm numbers, and the oracle encoder, used only as a
reference. The oracle matters because a gap in \auc{} is hard to interpret on its own. The drop from
oracle to causal reading gives that gap a scale, and we read every other gap against it. All
headline numbers use Whisper read at layer~6 (Section~\ref{sec:extra}), $W{=}10$, and a $5\%$
false-alarm budget.

\subsection{RQ1: Zero-shot transfer across languages}
\label{sec:e1}
\begin{figure}[t]
\centering
\includegraphics[width=\columnwidth]{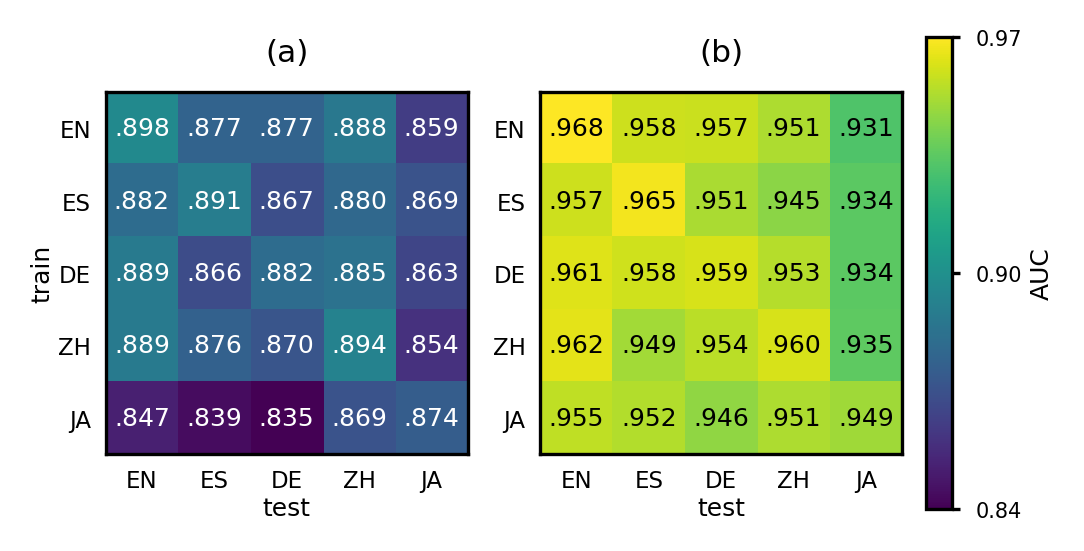}
\caption{Zero-shot transfer \auc{} for (a) causal and (b) full-context (oracle)
encoders (layer~6).}
\label{fig:e1}
\end{figure}

Reuse only helps if the frozen encoder represents turn ends in much the same way across languages,
so we begin by testing that. We train the head on each source language and evaluate it on every
target, which gives a full $5\times5$ source-by-target matrix for each encoder (Figure~\ref{fig:e1}).
The diagonal is the same-language ceiling and the off-diagonal is zero-shot transfer, and the
transfer gap $\Delta$ of Eq.~\eqref{eq:gap} is the average distance between the two. A head is worth
reusing only when this gap is small.

Transfer turns out to be strong, and it does not fall off from one language family to another. The
gap across the five languages is only $0.017$ \auc{} ($95\%$ CI $[0.016,0.019]$, positive in over
$99.9\%$ of resamples), so a head reads four unseen languages almost as well as their own native
heads do (Fig.~\ref{fig:e1}a). The oracle matrix shows the same even pattern at a slightly smaller
gap of $0.011$ \auc{} (Fig.~\ref{fig:e1}b). The oracle scale places this: moving from oracle to causal reading drops the diagonal by about
$0.07$ \auc{}, so going cross-lingual costs roughly a quarter of what going causal does.

These numbers describe the cleanly-yielded turn ends a detector is scored on, not every turn
transition. The share of ends dropped as active overlap ranges from about $0.40$ for English
and Mandarin to $0.48$, $0.53$, and $0.55$ for German, Spanish, and Japanese, so the
higher-overlap languages contribute a more selected set of events.

We had expected the tonal and pitch-accent languages to be the hard case, on the idea that a
language which uses pitch to distinguish words has less of it left to mark the end of a turn. The
results do not support that, since Mandarin transfers about as well as the non-tonal languages, and
the only language that is a little harder is Japanese. Even there the difficulty is in reading Japanese
as a target rather than in transferring from it, since foreign heads read it almost as well as its
own head does. Spending pitch on tone or accent therefore does not seem to weaken the cues a head uses to
find turn ends, since turn-final marking is distributed: final lengthening, decaying intensity, and contour shape all
mark the boundary, so the pitch a tonal language commits to word meaning is not the pitch
the boundary relies on. Because every cell uses the same fixed-capacity head, this is a like-for-like
comparison, and a higher-capacity nonlinear probe leaves the gap unchanged (Section~\ref{sec:extra}).

\textbf{Implication.} A single head covers the whole language set at a small fraction of the cost of
training one per language, and the encoder's streaming behavior matters more for quality than the
languages it was trained on.

\subsection{RQ2: Few-shot adaptation to a new language}
\label{sec:e2}
\begin{figure}[t]
\centering
\includegraphics[width=\columnwidth]{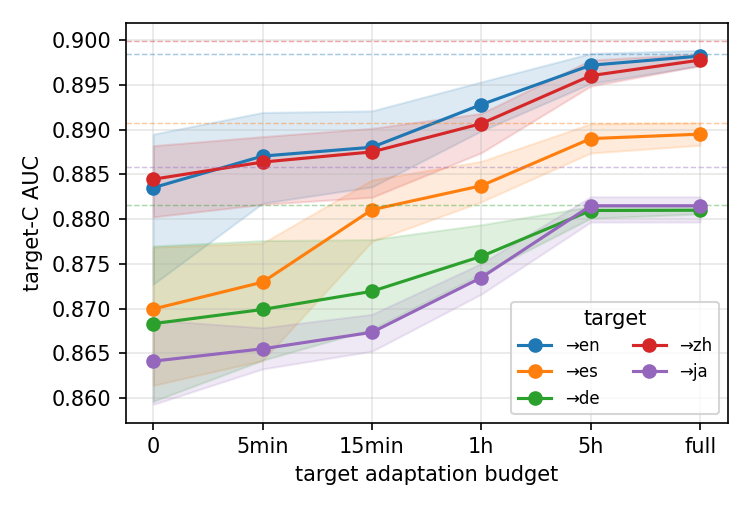}
\caption{Few-shot adaptation \auc{} versus target-data budget, per target language
(causal, layer~6).}
\label{fig:e2}
\end{figure}

A small gap is encouraging, but a small gap is not the same as one that closes cheaply, so the next
question is how much target-language audio it takes to remove what remains. We take each zero-shot
head and adapt it on a growing amount of target audio, from none up to the full adaptation split of
about six to seven hours, and track \auc{} against the same-language ceiling. Figure~\ref{fig:e2}
shows one curve for each target language, averaged over the four source languages, with bands
covering the range of sources and dashed lines marking the ceilings. The shape of each curve tells
us whether bringing up a new language is a light touch or a full rebuild.

The gap closes quickly at first and then levels off. Across all twenty directed pairs the zero-shot
gap starts small, between $0.005$ and $0.029$ \auc{}, and it shrinks steadily as more audio is
added. On the median pair, about $25\%$ of the gap is recovered within fifteen minutes, half
within an hour, and most of it within five hours, after which the curves flatten (Figure~\ref{fig:e2}).
Every target ends up within $0.004$ \auc{} of its own ceiling, including Japanese, which is the
hardest target and the one that starts with the largest gap. Since the gap stays under $0.03$ \auc{}
throughout, these fractions describe the shape of the curve rather than precise effect sizes.

\textbf{Implication.} Onboarding a new language is therefore light adaptation of an existing head,
not the collection of a full corpus, since a few hours of target audio recover almost all of the
remaining gap.

\subsection{RQ3: Stability of the operating point}
\label{sec:e3}
A head that transfers cheaply is still only useful if its threshold transfers too. If the threshold
had to be tuned again for each language, the per-language cost would simply reappear at the operating
point. We therefore take the transferable threshold of Section~\ref{sec:data}, set on the source
language and reused unchanged on the target, and measure what a real-time system would see, namely the
realized false-alarm rate and the endpoint latency. Both are reported for each target language in
Table~\ref{tab:e3}.

\begin{table}[t]
\centering
\caption{Operating-point transfer per target language when the source-calibrated
threshold $\theta$ is reused unchanged on the target (causal, layer~6). FA\,\% ranges
over the four source languages, and latency is p50/p90/p95 in ms.}
\label{tab:e3}
\begin{tabular}{lcc}
\toprule
Target & FA\,\% & lat p50/p90/p95 \\
\midrule
English  & 4.2--6.5 & $-80$/$40$/$80$ \\
Spanish  & 4.1--5.6 & $-80$/$20$/$60$ \\
German   & 3.5--4.9 & $-80$/$40$/$80$ \\
Mandarin & 4.1--5.6 & $-80$/$40$/$80$ \\
Japanese & 5.0--6.2 & $-80$/$20$/$60$ \\
\bottomrule
\end{tabular}
\end{table}

The operating point holds up well across all twenty directed pairs, where the false-alarm rate stays
between $3.5$ and $6.5\%$, within about a point and a half of the $5\%$ budget, and the median
latency moves by $0$ or $20$\,ms relative to a threshold retuned on the target (Table~\ref{tab:e3}).
On a typical turn the detector fires a little early, with a median of about $80$\,ms before the
boundary, and the late tail is short, with the 95th percentile no more than $80$\,ms after it. The
firing rate stays between $66$ and $82\%$ of turns across pairs, so it is the whole distribution that
carries over, not only the median.

\textbf{Implication.} Because a threshold set on one language keeps both its false-alarm budget and
its latency on another, the operating point travels with the head, and re-tuning it for each new
language becomes optional rather than necessary. JAL-Turn, by contrast, reports its results at a
single fixed decision threshold and does not test whether it holds across languages~\cite{jalturn2026},
so the cross-lingual behavior of the operating point has not been measured before.

\subsection{Robustness}
\label{sec:extra}
Taken together, the three results point to a portable detector. What follows asks whether that
portability is a real property of the representation or a by-product of the choices we fixed along
the way. The gap is insensitive to the two main labeling and modeling choices. Varying the collar
half-width moves it only from $0.015$ to $0.021$ \auc{} as $W$ goes from $5$ to $15$, and replacing
the logistic head with a calibrated single-hidden-layer probe leaves both the ordering of the cells
and the size of the gap unchanged. The checks below address the doubts that carry more weight, and
Table~\ref{tab:robust} collects the transfer gap for each configuration.

\begin{table}[t]
\centering
\caption{Five-language transfer gap $\Delta$ across encoder configurations (Whisper at layer~6
unless noted). Smaller is better.}
\label{tab:robust}
\begin{tabular}{lcc}
\toprule
Configuration & Causal $\Delta$ & Oracle $\Delta$ \\
\midrule
Whisper, frozen (main)            & 0.017 & 0.011 \\
Whisper, top-2 layers fine-tuned  & 0.023 & 0.008 \\
XLS-R, frozen                     & 0.009 & 0.012 \\
\bottomrule
\end{tabular}
\end{table}

\textbf{Could frame position alone explain the diagonal?} Because every example ends shortly after
the turn boundary, the positive collar always falls near the end of the window, so a probe could in
principle score well by reading position rather than acoustics. Weak timing-only references are the
usual way to rule this out~\cite{ekstedt2022vap}. A baseline given only each frame's relative
position, with no encoder features, scores $0.47$ to $0.59$ \auc{} across languages, against a
chance value of $0.5$. Position on its own carries almost no information, so the diagonal comes from
acoustic structure and not from where a frame sits in the window.

\textbf{Would fine-tuning the encoder do better?} Fine-tuning the top two encoder layers per source
language raises the mean same-language causal diagonal from $0.888$ to $0.906$, but widens the
causal transfer gap from $0.017$ to $0.023$, with non-overlapping $95\%$ CIs of $[0.016,0.019]$ and
$[0.021,0.025]$. With full context the gap narrows from $0.011$ to $0.008$, so the loss is confined
to causal reading. The direction matches what multilingual VAP found~\cite{inoue2024multilingual},
where a model trained per language transferred poorly while a single shared model did not, and it
appears here even though we adapt only two layers rather than retrain the encoder. Fine-tuning
raises same-language quality but specializes the representation to one language, whereas the frozen
encoder keeps the shared structure that transfer depends on.

\textbf{Is the result specific to Whisper?} A frozen XLS-R-300m encoder~\cite{babu2022xlsr} gives a
five-language gap of $0.009$ \auc{} causal and $0.012$ with full context, with German and Japanese
transferring like the rest, so the result holds for frozen multilingual encoders in general.
Read-out depth matters more here, which we check on the three languages of this comparison,
English, Spanish, and Mandarin. On those three, XLS-R's final layer collapses under causal
reading to $0.670$, far below Whisper's layer-6 read-out at $0.878$, but XLS-R's best mid-stack
layer, layer~$18$ of $24$, recovers to $0.861$ and matches it, so the read-out layer must be chosen
on causal \auc{} rather than the full-context score.

\begin{figure}[t]
\centering
\includegraphics[width=0.92\columnwidth]{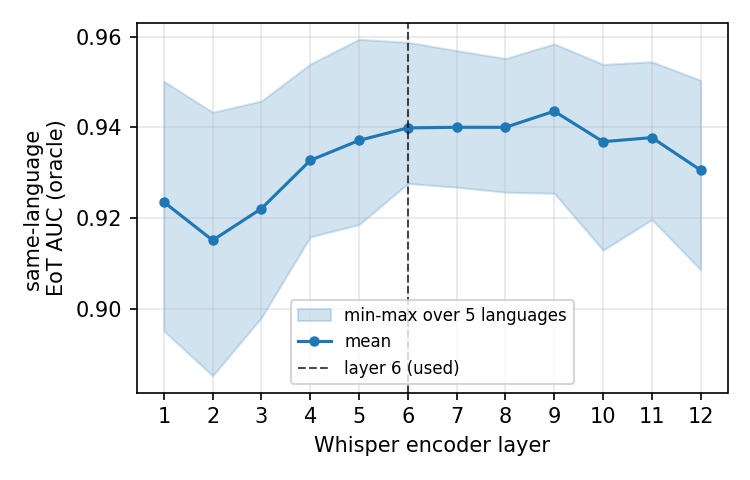}
\caption{Same-language EoT \auc{} versus Whisper encoder layer (oracle), the mean over the five
languages with a band covering their range. The dashed line marks the layer~6 read-out.}
\label{fig:layersweep}
\end{figure}

\textbf{Which layer should be read?} Across the twelve Whisper-small layers in oracle mode, the
information useful for EoT rises to a plateau over layers~$5$ to $9$ and then falls off at the final
layer (Figure~\ref{fig:layersweep}). The transfer gap is smallest over that same plateau. This matches
the top layer being the most ASR-specific and least prosodic~\cite{pasad2021layerwise,pasad2023comparative}.
We read layer~$6$ throughout, and any layer in the plateau gives a similar diagonal and gap.

Across these checks the gap stays near $0.01$, and only per-language fine-tuning widens it
(Table~\ref{tab:robust}).

\section{Limitations}
Our evidence comes from CALLHOME telephone speech, which is two-party and $8$\,kHz. Holding the
domain fixed is what makes the languages comparable, but it also means we have not shown that the
result carries to wideband, multi-party, or noisy far-field audio. The labels add a second
boundary, since we count a same-speaker resumption after a silence longer than $0.5$\,s as a turn
end, and a stricter threshold would move some of these holds back to within-turn pauses. We also score only
cleanly-yielded turn ends, excluding active overlaps, whose language-varying scope we bound in
Section~\ref{sec:data}. The frozen-versus-fine-tuned comparison is similarly bounded, since we
fine-tune only the top two encoder layers, so we show that light per-language adaptation already
costs transfer, not that freezing beats every possible adaptation regime. Finally, the study covers two
encoders, Whisper and XLS-R, and five languages from four families, which is a small sample of the
world's prosodic systems, so the honest reading is that EoT transfers well across the languages we
tested, with reasons to expect it holds more widely.

\section{Conclusion}
The cues that mark the end of a turn appear to live in a representation that is largely shared across
languages. A frozen multilingual encoder captures them well enough that a single head reads any of
our five languages, tonal Mandarin and pitch-accent Japanese included, almost as well as a native
head. Freezing preserves this shared structure and per-language fine-tuning erodes it, the same
pattern multilingual VAP reported~\cite{inoue2024multilingual}. Language coverage is therefore the
cheap part of an EoT detector and causal operation the expensive part, since going cross-lingual
costs about a quarter of what going causal does.

For a practitioner this points to a simple recipe, in which one head is trained on whatever language
is available, reused with its threshold across the rest, and supplemented with a few hours of target
audio only where a residual gap is worth closing, as for Japanese. The one part that does not transfer is the encoder and its read-out layer,
which must be chosen on causal rather than full-context \auc{}. Most of the per-language cost then
disappears, and how far this holds beyond telephone speech and five languages is the natural next
step.

\bibliographystyle{IEEEtran}
\bibliography{refs}

@inproceedings{baevski2020wav2vec2,
  author    = {Baevski, Alexei and Zhou, Henry and Mohamed, Abdelrahman and Auli, Michael},
  title     = {wav2vec 2.0: A Framework for Self-Supervised Learning of Speech
               Representations},
  booktitle = {Advances in Neural Information Processing Systems (NeurIPS)}, year = {2020},
  note      = {arXiv:2006.11477}}

@inproceedings{radford2023whisper,
  title     = {Robust Speech Recognition via Large-Scale Weak Supervision},
  author    = {Radford, Alec and Kim, Jong Wook and Xu, Tao and Brockman, Greg and McLeavey, Christine and Sutskever, Ilya},
  booktitle = {Proceedings of the 40th International Conference on Machine Learning (ICML)},
  series    = {Proceedings of Machine Learning Research},
  volume    = {202},
  pages     = {28492--28518},
  year      = {2023},
  publisher = {PMLR},
  url       = {https://proceedings.mlr.press/v202/radford23a.html}}

@inproceedings{ekstedt2022vap,
  title     = {Voice Activity Projection: Self-supervised Learning of Turn-taking Events},
  author    = {Ekstedt, Erik and Skantze, Gabriel},
  booktitle = {Proc. Interspeech}, year = {2022}, note = {arXiv:2205.09812}}

@inproceedings{ekstedt2022prosody,
  title     = {How Much Does Prosody Help Turn-taking? Investigations using Voice Activity
               Projection Models},
  author    = {Ekstedt, Erik and Skantze, Gabriel},
  booktitle = {Proc. SIGDIAL}, year = {2022}, note = {arXiv:2209.05161}}

@article{skantze2021review,
  title   = {Turn-taking in Conversational Systems and Human-Robot Interaction: A Review},
  author  = {Skantze, Gabriel},
  journal = {Computer Speech \& Language}, volume = {67}, pages = {101178}, year = {2021},
  doi     = {10.1016/j.csl.2020.101178}}

@inproceedings{pasad2021layerwise,
  title     = {Layer-wise Analysis of a Self-supervised Speech Representation Model},
  author    = {Pasad, Ankita and Chou, Ju-Chieh and Livescu, Karen},
  booktitle = {Proc. IEEE ASRU}, year = {2021}, note = {arXiv:2107.04734}}

@inproceedings{ferrer2002speaker,
  author    = {Ferrer, Luciana and Shriberg, Elizabeth and Stolcke, Andreas},
  title     = {Is the Speaker Done Yet? Faster and More Accurate End-of-Utterance Detection
               Using Prosody},
  booktitle = {Proc. 7th Int. Conf. on Spoken Language Processing (ICSLP)},
  pages     = {2061--2064}, year = {2002}, doi = {10.21437/ICSLP.2002-574}}

@inproceedings{ferrer2003prosody,
  author    = {Ferrer, Luciana and Shriberg, Elizabeth and Stolcke, Andreas},
  title     = {A Prosody-Based Approach to End-of-Utterance Detection That Does Not Require
               Speech Recognition},
  booktitle = {Proc. IEEE Int. Conf. on Acoustics, Speech and Signal Processing (ICASSP)},
  volume    = {1}, pages = {I-608--I-611}, year = {2003}, doi = {10.1109/ICASSP.2003.1198854}}

@article{gravano2011turn,
  title   = {Turn-taking cues in task-oriented dialogue},
  author  = {Gravano, Agust{\'i}n and Hirschberg, Julia},
  journal = {Computer Speech \& Language}, volume = {25}, number = {3}, pages = {601--634},
  year    = {2011}, doi = {10.1016/j.csl.2010.10.003}}

@article{sacks1974systematics,
  author  = {Sacks, Harvey and Schegloff, Emanuel A. and Jefferson, Gail},
  title   = {A Simplest Systematics for the Organization of Turn-Taking for Conversation},
  journal = {Language}, volume = {50}, number = {4}, pages = {696--735}, year = {1974},
  doi     = {10.2307/412243}}

@inproceedings{babu2022xlsr,
  title={{XLS-R}: Self-supervised Cross-lingual Speech Representation Learning at Scale},
  author={Babu, Arun and Wang, Changhan and Tjandra, Andros and Lakhotia, Kushal and Xu, Qiantong and Goyal, Naman and Singh, Kritika and von Platen, Patrick and Saraf, Yatharth and Pino, Juan and Baevski, Alexei and Conneau, Alexis and Auli, Michael},
  booktitle={Proc. Interspeech},
  year={2022},
  note={arXiv:2111.09296}}

@article{pratap2024mms,
  title={Scaling Speech Technology to 1,000+ Languages},
  author={Pratap, Vineel and Tjandra, Andros and Shi, Bowen and Tomasello, Paden and Babu, Arun and Kundu, Sayani and Elkahky, Ali and Ni, Zhaoheng and Vyas, Apoorv and Fazel-Zarandi, Maryam and Baevski, Alexei and Adi, Yossi and Zhang, Xiaohui and Hsu, Wei-Ning and Conneau, Alexis and Auli, Michael},
  journal={Journal of Machine Learning Research},
  volume={25}, year={2024}, note={arXiv:2305.13516}}

@inproceedings{shi2021emformer,
  title={{Emformer}: Efficient Memory Transformer Based Acoustic Model for Low Latency Streaming Speech Recognition},
  author={Shi, Yangyang and Wang, Yongqiang and Wu, Chunyang and Yeh, Ching-Feng and Chan, Julian and Zhang, Frank and Le, Duc and Seltzer, Mike},
  booktitle={Proc. IEEE ICASSP}, year={2021}, note={arXiv:2010.10759}}

@inproceedings{shi2023mlsuperb,
  title={{ML-SUPERB}: Multilingual Speech Universal PERformance Benchmark},
  author={Shi, Jiatong and Berrebbi, Dan and Chen, William and Chung, Ho-Lam and Hu, En-Pei and Huang, Wei Ping and Chang, Xuankai and Li, Shang-Wen and Mohamed, Abdelrahman and Lee, Hung-yi and Watanabe, Shinji},
  booktitle={Proc. Interspeech}, year={2023}, note={arXiv:2305.10615}}

@inproceedings{yang2021superb,
  title={{SUPERB}: Speech processing Universal PERformance Benchmark},
  author={Yang, Shu-wen and Chi, Po-Han and Chuang, Yung-Sung and Lai, Cheng-I Jeff and Lakhotia, Kushal and Lin, Yist Y. and Liu, Andy T. and Shi, Jiatong and Chang, Xuankai and Lin, Guan-Ting and others},
  booktitle={Proc. Interspeech}, year={2021}, note={arXiv:2105.01051}}

@article{alain2017probing,
  title={Understanding intermediate layers using linear classifier probes},
  author={Alain, Guillaume and Bengio, Yoshua},
  journal={arXiv preprint arXiv:1610.01644}, year={2017},
  note={ICLR 2017 Workshop}}

@article{efron1986bootstrap,
  title={Bootstrap Methods for Standard Errors, Confidence Intervals, and Other Measures of Statistical Accuracy},
  author={Efron, Bradley and Tibshirani, Robert},
  journal={Statistical Science}, volume={1}, number={1}, pages={54--75}, year={1986}}

@inproceedings{bisani2004bootstrap,
  title={Bootstrap estimates for confidence intervals in {ASR} performance evaluation},
  author={Bisani, Maximilian and Ney, Hermann},
  booktitle={IEEE International Conference on Acoustics, Speech, and Signal Processing (ICASSP)},
  volume={1}, pages={409--412}, year={2004}}

@inproceedings{koehn2004significance,
  title={Statistical significance tests for machine translation evaluation},
  author={Koehn, Philipp},
  booktitle={Proceedings of the 2004 Conference on Empirical Methods in Natural Language Processing (EMNLP)},
  pages={388--395}, year={2004}}

@article{stivers2009universals,
  title={Universals and cultural variation in turn-taking in conversation},
  author={Stivers, Tanya and Enfield, N. J. and Brown, Penelope and Englert, Christina and Hayashi, Makoto and Heinemann, Trine and Hoymann, Gertie and Rossano, Federico and de Ruiter, Jan Peter and Yoon, Kyung-Eun and Levinson, Stephen C.},
  journal={Proceedings of the National Academy of Sciences}, volume={106}, number={26}, pages={10587--10592}, year={2009}}

@article{levinson2015timing,
  title={Timing in turn-taking and its implications for processing language},
  author={Levinson, Stephen C. and Torreira, Francisco},
  journal={Frontiers in Psychology}, volume={6}, pages={731}, year={2015}}

@article{heldner2010,
  title={Pauses, gaps and overlaps in conversations},
  author={Heldner, Mattias and Edlund, Jens},
  journal={Journal of Phonetics}, volume={38}, number={4}, pages={555--568}, year={2010}}

@article{ward2000prosodic,
  title={Prosodic features which cue back-channel responses in {English} and {Japanese}},
  author={Ward, Nigel and Tsukahara, Wataru},
  journal={Journal of Pragmatics}, volume={32}, number={8}, pages={1177--1207}, year={2000}}

@inproceedings{inoue2024multilingual,
  title={Multilingual Turn-taking Prediction Using Voice Activity Projection},
  author={Inoue, Koji and Jiang, Bing'er and Ekstedt, Erik and Kawahara, Tatsuya and Skantze, Gabriel},
  booktitle={Proc. LREC-COLING}, year={2024}, note={arXiv:2403.06487}}

@inproceedings{inoue2024realtime,
  title={Real-time and Continuous Turn-taking Prediction Using Voice Activity Projection},
  author={Inoue, Koji and Jiang, Bing'er and Ekstedt, Erik and Kawahara, Tatsuya and Skantze, Gabriel},
  booktitle={Proc. Int. Workshop on Spoken Dialogue Systems Technology (IWSDS)}, year={2024},
  note={arXiv:2401.04868}}

@inproceedings{inoue2025backchannel,
  title={Yeah, Un, Oh: Continuous and Real-time Backchannel Prediction with Fine-tuning of Voice Activity Projection},
  author={Inoue, Koji and Jiang, Bing'er and Ekstedt, Erik and Kawahara, Tatsuya and Skantze, Gabriel},
  booktitle={Proc. NAACL-HLT}, year={2025}, note={arXiv:2410.15929}}

@inproceedings{chang2022turn,
  title={Turn-Taking Prediction for Natural Conversational Speech},
  author={Chang, Shuo-yiin and Li, Bo and Sainath, Tara N. and Zhang, Chao and Strohman, Trevor and Liang, Qiao and He, Yanzhang},
  booktitle={Proc. Interspeech}, year={2022}, note={arXiv:2208.13321}}

@inproceedings{shangguan2021dissecting,
  title={Dissecting User-Perceived Latency of On-Device {E2E} Speech Recognition},
  author={Shangguan, Yuan and Prabhavalkar, Rohit and Su, Hang and Mahadeokar, Jay and Shi, Yangyang and Zhou, Jiatong and Wu, Chunyang and Le, Duc and Kalinli, Ozlem and Fuegen, Christian and Seltzer, Michael L.},
  booktitle={Proc. Interspeech}, year={2021}, note={arXiv:2104.02207}}

@inproceedings{arsikere2014enhanced,
  title={Computationally-efficient endpointing features for natural spoken interaction with personal-assistant systems},
  author={Arsikere, Harish and Shriberg, Elizabeth and Ozertem, Umut},
  booktitle={Proc. IEEE ICASSP}, year={2014}, doi={10.1109/ICASSP.2014.6854199}}

@misc{ldc1997callhome,
  title={{CALLHOME} Multilingual Telephone Speech Corpora},
  author={{Linguistic Data Consortium}},
  year={1996--1997},
  note={LDC catalog: CALLHOME American English (LDC97S42), Spanish (LDC96S35), German (LDC97S43), Mandarin (LDC96S34), Japanese (LDC96S37)}}

@misc{silero2021vad,
  title={{Silero VAD}: pre-trained enterprise-grade Voice Activity Detector},
  author={{Silero Team}}, year={2021},
  howpublished={\url{https://github.com/snakers4/silero-vad}}, note={Accessed 2026}}

@misc{livekit2024turn,
  title        = {Using a Transformer to Improve End-of-Turn Detection},
  author       = {{LiveKit}}, year = {2024},
  howpublished = {\url{https://livekit.com/blog/using-a-transformer-to-improve-end-of-turn-detection}},
  note         = {Blog post. 135M-parameter {SmolLM} v2 transcript-based end-of-utterance model. Accessed 2026}}

@misc{pipecat2025smartturn,
  title        = {Smart Turn Detection},
  author       = {{Pipecat AI}}, year = {2025},
  howpublished = {\url{https://docs.pipecat.ai/server/utilities/smart-turn/smart-turn-overview}},
  note         = {Product documentation. Smart Turn v3, a Whisper Tiny-based model with a small classifier head. Accessed 2026}}

@inproceedings{tan2020crosslingual,
  title={Evaluating Cross-Lingual Transfer Learning Approaches in Multilingual Conversational Agent Models},
  author={Tan, Lizhen and Golovneva, Olga},
  booktitle={Proc. 28th Int. Conf. on Computational Linguistics: Industry Track (COLING)},
  pages={1--9}, year={2020},
  doi={10.18653/v1/2020.coling-industry.1}}

@inproceedings{maas2018eou,
  title={Combining Acoustic Embeddings and Decoding Features for End-of-Utterance Detection in Real-Time Far-Field Speech Recognition Systems},
  author={Maas, Roland and Rastrow, Ariya and Ma, Chengyuan and Lan, Guitang and Goehner, Kyle and Tiwari, Gautam and Joseph, Shaun and Hoffmeister, Bj\"orn},
  booktitle={Proc. IEEE Int. Conf. on Acoustics, Speech and Signal Processing (ICASSP)},
  pages={5544--5548}, year={2018},
  doi={10.1109/ICASSP.2018.8461478}}

@misc{assemblyai2025endpointing,
  title        = {How Intelligent Turn Detection (Endpointing) Solves the Biggest Challenge in Voice Agent Development},
  author       = {{AssemblyAI}}, year = {2025},
  howpublished = {\url{https://www.assemblyai.com/blog/turn-detection-endpointing-voice-agent}},
  note         = {Blog post. Semantic endpointing in {Universal-Streaming} via special-token prediction. Accessed 2026}}

@book{yip2002tone,
  title={Tone},
  author={Yip, Moira},
  year={2002},
  publisher={Cambridge University Press},
  series={Cambridge Textbooks in Linguistics}}

@article{beckman1986japanese,
  title={Intonational structure in {Japanese} and {English}},
  author={Beckman, Mary E. and Pierrehumbert, Janet B.},
  journal={Phonology Yearbook},
  volume={3}, pages={255--309}, year={1986},
  doi={10.1017/S095267570000066X}}

@book{ladd2008intonational,
  title={Intonational Phonology},
  author={Ladd, D. Robert},
  edition={2}, year={2008},
  publisher={Cambridge University Press}}

@misc{glottolog2023,
  title={Glottolog 4.8},
  author={Hammarstr{\"o}m, Harald and Forkel, Robert and Haspelmath, Martin and Bank, Sebastian},
  year={2023},
  publisher={Max Planck Institute for Evolutionary Anthropology},
  address={Leipzig},
  howpublished={\url{https://glottolog.org}}, note={Accessed 2026}}

@article{bogels2015intonation,
  title={Listeners use intonational phrase boundaries to project turn ends in spoken interaction},
  author={B{\"o}gels, Sara and Torreira, Francisco},
  journal={Journal of Phonetics},
  volume={52}, pages={46--57}, year={2015},
  doi={10.1016/j.wocn.2015.04.004}}

@article{deruiter2006projecting,
  title   = {Projecting the End of a Speaker's Turn: A Cognitive Cornerstone of Conversation},
  author  = {de Ruiter, Jan P. and Mitterer, Holger and Enfield, N. J.},
  journal = {Language}, volume = {82}, number = {3}, pages = {515--535}, year = {2006},
  doi     = {10.1353/lan.2006.0130}}

@inproceedings{skantze2017continuous,
  title     = {Towards a General, Continuous Model of Turn-taking in Spoken Dialogue using {LSTM} Recurrent Neural Networks},
  author    = {Skantze, Gabriel},
  booktitle = {Proc. 18th Annual SIGdial Meeting on Discourse and Dialogue},
  pages     = {220--230}, year = {2017}, doi = {10.18653/v1/W17-5527}}

@inproceedings{roddy2018investigating,
  title     = {Investigating Speech Features for Continuous Turn-Taking Prediction Using {LSTMs}},
  author    = {Roddy, Matthew and Skantze, Gabriel and Harte, Naomi},
  booktitle = {Proc. Interspeech}, year = {2018}, note = {arXiv:1806.11461}}

@inproceedings{chang2017endpoint,
  title     = {Endpoint Detection Using Grid Long Short-Term Memory Networks for Streaming Speech Recognition},
  author    = {Chang, Shuo-Yiin and Li, Bo and Sainath, Tara N. and Simko, Gabor and Parada, Carolina},
  booktitle = {Proc. Interspeech}, pages = {3812--3816}, year = {2017},
  doi       = {10.21437/Interspeech.2017-284}}

@inproceedings{li2020streaming,
  title     = {Towards Fast and Accurate Streaming End-to-End {ASR}},
  author    = {Li, Bo and Chang, Shuo-Yiin and Sainath, Tara N. and Pang, Ruoming and He, Yanzhang and Strohman, Trevor and Wu, Yonghui},
  booktitle = {Proc. IEEE ICASSP}, year = {2020}, note = {arXiv:2004.11544}}

@inproceedings{bijwadia2022unified,
  title     = {Unified End-to-End Speech Recognition and Endpointing for Fast and Efficient Speech Systems},
  author    = {Bijwadia, Shaan and Chang, Shuo-Yiin and Li, Bo and Sainath, Tara N. and Zhang, Chao and He, Yanzhang},
  booktitle = {Proc. IEEE Spoken Language Technology Workshop (SLT)}, year = {2022}, note = {arXiv:2211.00786}}

@inproceedings{aldeneh2018endofturn,
  title     = {Improving End-of-Turn Detection in Spoken Dialogues by Detecting Speaker Intentions as a Secondary Task},
  author    = {Aldeneh, Zakaria and Dimitriadis, Dimitrios and Mower Provost, Emily},
  booktitle = {Proc. IEEE ICASSP}, year = {2018}, doi = {10.1109/ICASSP.2018.8461997}}

@misc{popit2025thai,
  title  = {Thai Semantic End-of-Turn Detection for Real-Time Voice Agents},
  author = {Popit, Thanapol and others},
  year   = {2025}, note = {arXiv:2510.04016}}

@article{hsu2021hubert,
  title   = {{HuBERT}: Self-Supervised Speech Representation Learning by Masked Prediction of Hidden Units},
  author  = {Hsu, Wei-Ning and Bolte, Benjamin and Tsai, Yao-Hung Hubert and Lakhotia, Kushal and Salakhutdinov, Ruslan and Mohamed, Abdelrahman},
  journal = {IEEE/ACM Transactions on Audio, Speech, and Language Processing},
  volume  = {29}, pages = {3451--3460}, year = {2021}, note = {arXiv:2106.07447}}

@article{chen2022wavlm,
  title   = {{WavLM}: Large-Scale Self-Supervised Pre-Training for Full Stack Speech Processing},
  author  = {Chen, Sanyuan and Wang, Chengyi and Chen, Zhengyang and Wu, Yu and Liu, Shujie and Chen, Zhuo and Li, Jinyu and Kanda, Naoyuki and Yoshioka, Takuya and Xiao, Xiong and Wu, Jian and Zhou, Long and Ren, Shuo and Qian, Yanmin and Qian, Yao and Zeng, Michael and Wei, Furu},
  journal = {IEEE Journal of Selected Topics in Signal Processing}, volume = {16}, number = {6},
  pages   = {1505--1518}, year = {2022}, doi = {10.1109/JSTSP.2022.3188113}}

@inproceedings{conneau2021xlsr,
  title     = {Unsupervised Cross-lingual Representation Learning for Speech Recognition},
  author    = {Conneau, Alexis and Baevski, Alexei and Collobert, Ronan and Mohamed, Abdelrahman and Auli, Michael},
  booktitle = {Proc. Interspeech}, year = {2021}, note = {arXiv:2006.13979}}

@inproceedings{pasad2023comparative,
  title     = {Comparative Layer-wise Analysis of Self-supervised Speech Models},
  author    = {Pasad, Ankita and Shi, Bowen and Livescu, Karen},
  booktitle = {Proc. IEEE ICASSP}, year = {2023}, note = {arXiv:2211.03929}}

@article{fawcett2006roc,
  title   = {An Introduction to {ROC} Analysis},
  author  = {Fawcett, Tom},
  journal = {Pattern Recognition Letters}, volume = {27}, number = {8}, pages = {861--874},
  year    = {2006}, doi = {10.1016/j.patrec.2005.10.010}}

@article{wightman1992segmental,
  title   = {Segmental Durations in the Vicinity of Prosodic Phrase Boundaries},
  author  = {Wightman, Colin W. and Shattuck-Hufnagel, Stefanie and Ostendorf, Mari and Price, Patti J.},
  journal = {The Journal of the Acoustical Society of America}, volume = {91}, number = {3},
  pages   = {1707--1717}, year = {1992}, doi = {10.1121/1.402450}}

@misc{chen2025fullduplex,
  title  = {From Turn-Taking to Synchronous Dialogue: A Survey of Full-Duplex Spoken Language Models},
  author = {Chen, Yuxuan and Yu, Haoyuan},
  year   = {2025}, note = {arXiv:2509.14515}}

@inproceedings{ekstedt2020turngpt,
  title     = {{TurnGPT}: A Transformer-based Language Model for Predicting Turn-taking in Spoken Dialog},
  author    = {Ekstedt, Erik and Skantze, Gabriel},
  booktitle = {Findings of the Association for Computational Linguistics: EMNLP 2020},
  year      = {2020}, doi = {10.18653/v1/2020.findings-emnlp.268}, note = {arXiv:2010.10874}}

@inproceedings{wang2024fusion,
  title     = {Turn-taking and Backchannel Prediction with Acoustic and Large Language Model Fusion},
  author    = {Wang, Jinhan and Chen, Long and Khare, Aparna and Raju, Anirudh and Dheram, Pranav and He, Di and Wu, Minhua and Stolcke, Andreas and Ravichandran, Venkatesh},
  booktitle = {Proc. IEEE ICASSP}, pages = {12121--12125}, year = {2024},
  doi = {10.1109/ICASSP48485.2024.10447196}}

@misc{jalturn2026,
  title  = {{JAL-Turn}: Joint Acoustic-Linguistic Modeling for Real-Time and Robust Turn-Taking Detection in Full-Duplex Spoken Dialogue Systems},
  author = {Yang, Guangzhao and Pan, Yu and Qiu, Shi and Bai, Ningjie},
  year   = {2026}, note = {arXiv:2603.26515}}

@misc{speechmatics2025semantic,
  title        = {End of Turn Detection},
  author       = {{Speechmatics}}, year = {2025},
  howpublished = {\url{https://docs.speechmatics.com/speech-to-text/realtime/end-of-turn}},
  note         = {Product documentation. Silence-based End of Utterance plus context-aware Semantic End of Turn Detection. Accessed 2026}}

@misc{assemblyai2025multilingual,
  title        = {Introducing Multilingual {Universal-Streaming}},
  author       = {{AssemblyAI}}, year = {2025},
  howpublished = {\url{https://www.assemblyai.com/blog/introducing-multilingual-universal-streaming}},
  note         = {Blog post. Multilingual real-time STT with built-in end-of-turn detection. Accessed 2026}}

@misc{livekit2025intl,
  title        = {Improved End-of-Turn Model Cuts Voice {AI} Interruptions 39\%},
  author       = {{LiveKit}}, year = {2025},
  howpublished = {\url{https://livekit.com/blog/improved-end-of-turn-model-cuts-voice-ai-interruptions-39}},
  note         = {Blog post. Multilingual model {v0.4.1-intl}, 39.23\% relative reduction in false-positive interruptions. Accessed 2026}}

@misc{livekit2026v1,
  title        = {Solving End-of-Turn Detection: {LiveKit} Turn Detector v1.0},
  author       = {{LiveKit}}, year = {2026},
  howpublished = {\url{https://livekit.com/blog/solving-end-of-turn-detection}},
  note         = {Blog post. Audio-based model fusing semantic and acoustic cues across 14 languages. Accessed 2026}}

@misc{russell2026prosodic,
  title  = {The Role of Prosodic and Lexical Cues in Turn-Taking with Self-Supervised Speech Representations},
  author = {{O'Connor Russell}, Sam and Charuau, Delphine and Harte, Naomi},
  year   = {2026}, note = {arXiv:2601.13835}}

@misc{udupa2026endpoint,
  title  = {Endpoint Anticipation for Low-Latency Spoken Dialogue},
  author = {Udupa, Sathvik and Watanabe, Shinji and Schwarz, Petr and {\v{C}}ernock{\'{y}}, Jan},
  year   = {2026}, note = {arXiv:2606.13450}}

@misc{helwani2026hierarchical,
  title  = {A Hierarchical End-of-Turn Model with Primary Speaker Segmentation for Real-Time Conversational {AI}},
  author = {Helwani, Karim and Do, Hoang and Luan, James and Srinivasan, Sriram},
  year   = {2026}, note = {arXiv:2603.13379}}


\end{document}